\documentclass[a4paper,12pt]{revtex4}
\usepackage{amsmath}  
\usepackage{amssymb}  
\usepackage{dsfont}   

\begin{document}
\title{Locus model for space-time fabric and quantum indeterminacies}
\author{Alberto C. de la Torre}\email{delatorre@mdp.edu.ar}
\address{ Universidad Nacional de Mar del Plata\\
  Argentina}
\begin{abstract}
A simple \emph{locus} model for the space-time fabric is presented
and is compared with quantum foam and random walk models. The
induced indeterminacies in momentum are calculated and it is shown
that these space-time fabric indeterminacies are, in most cases,
negligible compared with the quantum mechanical indeterminacies.
This result restricts the possibilities of an experimental
observation of the space-time fabric.
\\
Keywords: atomic space-time, Planck scale, indeterminacies.
\end{abstract}
\maketitle
\section{INTRODUCTION}
In this work a very simple atomic or discrete space-time model is
presented. The atoms of space-time, that we name \emph{loci}
(\emph{locus} in singular), have sizes comparable with Planck scale
and are located in a mathematical continuous space. All points
within a given \emph{locus} are physically equivalent and can not be
differentiated or taken apart.

We can consider these \emph{loci} as probability distributions for
physical coordinates and therefore distances, time intervals and
momenta become random variables that can be calculated from the
\emph{loci} distributions. We will find the probability uncertainty
for these quantities, without any reference to quantum mechanics,
and we will afterwards compare them with the corresponding quantum
mechanic indeterminacies.
\section{THE LOCUS MODEL}
Let us use the continuous real variables $(x,t)$ to denote the
\emph{mathematical coordinates} of a space-time point. We will
differentiate these mathematical coordinates from the \emph{physical
coordinates} because two points $x_{1}$ and $x_{2}$ separated by a
distance comparable with the Planck length $\ell_{p}$ are physically
indistinguishable and two instants of time separated by an interval
comparable with Planck time $t_{p}$ can not be considered as
physically different. In order to formalize this concept we propose
that a physical coordinate is described by space-time region, a
\emph{locus}, centered at a mathematical coordinate and having a
width given by the Planck scale. A space-time localization of a
particle means that a certain \emph{locus} is occupied and a
physical space-time interval will be determined by the set of
\emph{loci} between two space-time points. The precise shape and
boundaries of these \emph{loci} will not be relevant and we can also
imagine soft boundaries that could be described by a probability
density (gaussian for instance). In this way, physical coordinates
become random variables distributed with probability densities
$L_{x,\ell_{p}}(\xi)$ and $T_{t,t_{p}}(\theta)$ with center at
$(x,t)$ and widths $(\ell_{p},t_{p})$. The physically relevant
interval, or distance between two coordinates $x_{1}$ and $x_{2}$,
is then a random variable $x_{2}-x_{1}$ distributed according to the
convolution
\begin{equation}\label{conv}
    \int_{\infty}^{\infty} d\eta\  L_{x_{2},\ell_{p}}(\xi-\eta)L_{x_{1},\ell_{p}}(\eta)\
    .
\end{equation}
Let us consider now the physical space between two distant
coordinate points, say $x=0$ and $x=\ell$. We can define a partition
of the interval $0<x_{1}<x_{2},\cdots,<x_{N}<\ell$ and we have
$\ell=\ell-x_{N}+x_{N}-x_{N-1}+x_{N-1}-\cdots+x_{2}-x_{1}+x_{1}-0$.
So we have decomposed the interval $\ell$ in a sum of $N+1$
subintervals and therefore this physical length is a random variable
distributed as an $N+1$ fold convolution. If $N$ is large, the
distribution will approach a gaussian distribution, regardless of
the shape of the \emph{locus} distribution, with a width given by
$\sqrt{2N-2}\ell_{p}$. Accordingly, the distribution of a physical
length $\ell$ will depend on the number of points in the partitions.
We can fix the number of subintervals to be approximatively equal to
the number of \emph{loci} fitting in the length $\ell$. Let us
define then $\delta_{L}$ to be the space density of \emph{loci} and
then we have $N=\delta_{L}\ell$. We can expect that the space
density of \emph{loci} is close to $1/\ell_{p}$ because if the
density would be much larger, then we could have physical locations
separated by a distance less then $\ell_{p}$ and if it were much
smaller the transition from one location to the next would not be
possible. A physical length $\ell$, much longer than Planck length
$\ell_{p}$, is then a random variable with a gaussian distribution
\begin{equation}\label{ldist}
  \Xi(\xi)=\frac{1}{\sqrt{2\pi}\sigma_{x}}
    \exp\left(-\frac{(\xi-\ell)^{2}}{2\sigma_{x}^{2}}\right)\
\end{equation}
peaked at $\xi=\ell$ and with a width $\sigma_{x}$ (we reserve
$\Delta_{x}$ for quantum indeterminacies)
\begin{equation}\label{widthx}
    \sigma_{x}=\sqrt{2\ell\delta_{L}}\ell_{p}\ ,
\end{equation}
and if we take $\delta_{L}\approx 1/\ell_{p}$ we
have
\begin{equation}\label{widthx1}
    \sigma_{x}=\left(2\frac{\ell}{\ell_{p}}\right)^{1/2}\ell_{p}\ .
\end{equation}

In a similar way we conclude that a time interval $t$, much
longer than Planck time $t_{p}$, is a random variable with a
gaussian distribution
\begin{equation}\label{tdist}
 \Theta(\theta)=\frac{1}{\sqrt{2\pi}\sigma_{t}}
    \exp\left(-\frac{(\theta-t)^{2}}{2\sigma_{t}^{2}}\right)\
\end{equation}
with a width $\sigma_{t}$
\begin{equation}\label{widtht}
    \sigma_{t}=\sqrt{2t\delta_{T}}t_{p}\ ,
\end{equation}
where $\delta_{T}$ is the \emph{loci} time density and if we take it
$\delta_{T}\approx 1/t_{p}$ we have
\begin{equation}\label{widtht1}
    \sigma_{t}=\left(2\frac{t}{t_{p}}\right)^{1/2}t_{p}\ .
\end{equation}

We can now compare these results with other models with an
essential indeterminacy in space-time points. An early proposal
was made by Karolyhazy\cite{karo} that combined Heisenberg's
uncertainty principle with Schwarzschild horizon in order to
estimate a minimal uncertainty in the measurement of a distance
$\ell$ and a time $t$ given by
\begin{equation} \label{karol}
\sigma_{x}\sim\left(\frac{\ell}{\ell_{p}}\right)^{1/3}\ell_{p}
 \ \mbox{  , and }\
 \sigma_{t}\sim\left(\frac{t}{t_{p}}\right)^{1/3}t_{p}\ .
\end{equation}
Due to the $1/3$ exponent, these indeterminacies are much smaller
than the ones resulting from the \emph{locus} model. These results,
obtained from a heuristic argument, were rediscovered in the context
of a \emph{quantum foam} model for the space-time fabric\cite{ng1}
and they gain support from other apparently independent
arguments\cite{ng2}. Indeed it was shown that the same result can be
obtained as a consequence of the Holographic Principle and also from
Black Holes physics and even from Information and Computer Theory.
The \emph{locus} model result, with an $1/2$ exponent, was also
obtained from a random walk\cite{rw1,rw2} model. An interesting
feature of these random walk models is that they can be modified in
order to obtain also the $1/3$ exponent by the introduction of some
memory in the random walk that increases the probability of
returning to the previous position like a repentant walker. One
could motivate such a memory by an attractive self interaction
between the actual and the previous position of a particle.
\section{INDUCED INDETERMINACIES IN MOMENTUM}
In this section we will deduce the momentum indeterminacy induced by
the space-time indeterminacies. Although we will concentrate on the
indeterminacies of the \emph{locus} model, the conclusions are also
valid for  the other models with a $1/3$ exponent. Let us consider a
free particle of mass $m$ moving a distance $\ell$ during a time
interval $t$. Since these quantities are random variables with
distributions given in Eqs.(\ref{ldist}, \ref{tdist}) the momentum
of the particle, given by $p=m\ell/t$, will also be a random
variable with the distribution corresponding to the quotient of
random variables. Therefore the momentum $p$ will be distributed
according to the probability density function $\Pi(\varpi)$ given by
\begin{equation} \label{pdist}
 \Pi(\varpi)=\int_{-\infty}^{\infty}\!\!\!  d\xi
 \int_{-\infty}^{\infty}\!\!\!  d\theta\ \Xi(\xi)\ \Theta(\theta)\
 \delta\!\!\left(\varpi-m\frac{\xi}{\theta}\right)
 =\int_{-\infty}^{\infty}\!\!\!  d\theta\
 \left|\frac{\theta}{m}\right|\ \Xi(\varpi\theta)\
 \Theta(\theta)\ .
\end{equation}
If we insert the gaussian densities given in
Eqs.(\ref{ldist},\ref{tdist}) we can obtain the momentum density
distribution in terms of the Error Function and considering that
$t\gg t_{p}$ we get the approximation
\begin{equation}\label{pdist1}
   \Pi(\varpi)\approx\frac{\sqrt{pmc}}{2\pi}\
   \frac{\varpi +mc}{(\varpi^{2}+pmc)^{3/2}}
   \left(\frac{\ell}{\ell_{p}}\right)^{1/2}
   \exp\left(-\frac{(\varpi-p)^{2}}
   {4\frac{p}{mc}\frac{\ell_{p}}{\ell}(\varpi^{2}+pmc)}\right)\ .
\end{equation}
This distribution is peaked at $\varpi=p=m\ell/t$ and its width
can be estimated from the denominator of the exponent. However we
can obtain the momentum indeterminacy more rigorously from the
definition
\begin{eqnarray}
\nonumber
 \sigma_{p}^{2}&=& \int_{-\infty}^{\infty}\!\!\!  d\varpi\
  (\varpi-p)^{2}\ \Pi(\varpi) \\ \nonumber
 &=& \int_{-\infty}^{\infty}\!\!\!  d\varpi
  \int_{-\infty}^{\infty}\!\!\!  d\xi
 \int_{-\infty}^{\infty}\!\!\!  d\theta
 \ (\varpi-p)^{2}\ \Xi(\xi)\ \Theta(\theta)\
 \delta\!\!\left(\varpi-m\frac{\xi}{\theta}\right) \\ \nonumber
 &=&  \int_{-\infty}^{\infty}\!\!\!  d\xi
 \int_{-\infty}^{\infty}\!\!\!  d\theta\
 (m\frac{\xi}{\theta}-p)^{2}\ \Xi(\xi)\ \Theta(\theta)\ \\ \nonumber
 &=& m^{2} \int_{-\infty}^{\infty}\!\!\!  d\xi\ \xi^{2}\ \Xi(\xi)
 \int_{-\infty}^{\infty}\!\!\!  d\theta
 \ \frac{1}{\theta^{2}}\ \Theta(\theta)\ -\
2mp \int_{-\infty}^{\infty}\!\!\!  d\xi\ \xi\ \Xi(\xi)
 \int_{-\infty}^{\infty}\!\!\!  d\theta
 \ \frac{1}{\theta}\ \Theta(\theta) +p^{2} \\ \nonumber
&=& m^{2}(\sigma_{x}^{2}+\ell^{2})
\left\langle\frac{1}{\theta^{2}}\right\rangle
-2mp\ell\left\langle\frac{1}{\theta}\right\rangle + p^{2} \\
&=& m^{2}\sigma_{x}^{2}
\left\langle\frac{1}{\theta^{2}}\right\rangle
+m^{2}\ell^{2}\left(\left\langle\frac{1}{\theta^{2}}\right\rangle
-2\frac{1}{t}\left\langle\frac{1}{\theta}\right\rangle
+\frac{1}{t^{2}}\right)\ .
\end{eqnarray}
Since the $\Theta(\theta)$ distribution is sharply peaked at
$\theta=t$ we can take as good approximation $\langle
1/\theta^{2}\rangle=1/t^{2}$ and $\langle1/\theta\rangle=1/t$ and
with this, the parenthesis in last equation vanishes. With this
we become then the simple expression
\begin{equation}\label{sigxp}
  \frac{\sigma_{p}}{p}=\frac{\sigma_{x}}{\ell}\ .
\end{equation}
This is a general result but if we specialize it for the
\emph{locus} model, using Eqs.(\ref{widthx1} and \ref{widtht1}), we
obtain
\begin{equation}\label{sigxpt}
\frac{\sigma_{p}}{p}=\frac{\sigma_{x}}{\ell}=
\frac{\sigma_{t}}{t}\left(\frac{mc}{p}\right)^{1/2}=
\left(2\frac{\ell_{p}}{\ell}\right)^{1/2}\ .
\end{equation}
\section{COMPARISON WITH QUANTUM INDETERMINACIES}
In the estimation of the indeterminacies due to the space-time
fabric with a $1/3$ exponent, some quantum mechanical arguments
have been used. However, since these arguments were heuristic,
there is no guaranty that the indeterminacies obtained are
compatible with rigourous quantum mechanical indeterminacies and
their correlations manifest in the uncertainty principle. Quantum
mechanical indeterminacies are ubiquitous but nonintuitive and it
is therefore dangerous to identify them with indeterminacies
arising in idealized measurement procedures. Indeed, it is well
known that some heuristic arguments using Heisenberg's principle
may lead to erroneous results\cite{BohrEins} and as D. Griffith
warns ``when you hear a physicist invoke the uncertainty
principle, keep a hand on your wallet''\cite{grif}. Furthermore
the $1/2$ exponent indeterminacies are larger than the $1/3$ ones
and were derived without reference to quantum mechanics. It is
therefore necessary to compare the space-time fabric
indeterminacies $\sigma_{x}$ and $\sigma_{p}$ with the quantum
mechanical indeterminacies $\Delta_{x}$ and $\Delta_{p}$, in
particular with their correlations
$\Delta_{x}\Delta_{p}\geq\hbar/2$. From Eq.(\ref{sigxpt}) above
we immediately obtain for the product of the space-time fabric
indeterminacies
\begin{equation}\label{prodsxsp}
   \sigma_{x}\sigma_{p}=2p\ell_{p}\ ,
\end{equation}
and therefore for a momentum smaller than some value we have a
product of indeterminacies below the quantum mechanical bound
$\hbar/2$. This value of momentum turns out to be enormous, 5 kg m/s
or $10^{28}$ ev/c, many orders of magnitude bigger than the highest
energy cosmic rays observed.  We must conclude that the
indeterminacies due to the space-time fabric lie deep below the
quantum mechanical indeterminacies. This places severe limits on the
observability of the space-time indeterminacies in the kinematics of
a particle because any observation will encounter the quantum
mechanical limits long before the space-time indeterminacies are
approached. There are proposals\cite{ng3} to observe space-time
indeterminacies in extragalactic light by interferometric techniques
taking $\ell$ large enough, close to the observable universe length
$\ell\sim 10^{26}$m. In such a long travel, the photons in the
\emph{locus} or random walk models would develop an indeterminacy
$\sigma_{x}\sim 1.7\ 10^{-3}$m, much longer than the wave length,
making the light incoherent and therefore no interference should be
observed. In the case of the other models, the indeterminacy
accumulated is $\sigma_{x}\sim 3\ 10^{-15}$m, not sufficient to
destroy coherence. The observation of interference fringes seem to
exclude the $1/2$ exponent models. The observation of the space-time
indeterminacies for these $1/3$ exponent models faces the
difficulties imposed by quantum mechanics: in order to test an
indeterminacy of $\sigma_{x}\sim 3\ 10^{-15}$m the quantum
mechanical indeterminacy must be even smaller
$\Delta_{x}<\sigma_{x}$ but this implies a momentum indeterminacy
$\Delta_{p}>\hbar/(2 \sigma_{x})$ and this turns out to be 30 Mev/c
requiring very high energy gamma rays.
\section{CONCLUSIONS}
The \emph{locus} model presented is very simple and intuitive,
however some modifications and refinements could be necessary when
the observations of space-time indeterminacies becomes feasible.
Several possible changes in the model can be implemented. One of
them could to modify the space and time \emph{loci} densities
$\delta_{L}$ and $\delta_{T}$ considering the possibility of very
high or very low \emph{loci} densities or overlap. Another
modification could be to consider physical space and time
coordinates not as independent random variables but instead
described by a joint probability distribution function. There are
also more speculative possibilities where we could describe the
propagation in terms of a deformation of the \emph{loci} in the
direction of propagation by an amount related to the energy
enclosed. As is also the case for the other quantum foam models, the
\emph{locus} model violates Lorentz covariance and we could assume
the \emph{loci} to be at rest in the reference frame where the 2.7K
background radiation is isotropic, or to assume that they move with
a velocity distribution with a large spread in order to recover
approximatively Lorentz covariance.


\begin{thebibliography}{99}

\bibitem{karo}F. Karolyhazy, Nuovo Cimento \textbf{42}, 390
(1966).

\bibitem{ng1} Y. J. Ng and H. van Dam, Mod Phys. Lett \textbf{A9}, 335 (1994).

\bibitem{ng2} Y. J. Ng, Quantum Foam. arXiv:grt-qc/0401015.

\bibitem{rw1} L. Diosi and B. Lukacs, Phys. Lett.  \textbf{A42}, 331 (1989).

\bibitem{rw2} G. Amelini-Camelia, Nature \textbf{398}, 216 (1999).

\bibitem{BohrEins} A. C. de la Torre, A. Daleo and I. Garc\'{\i}a-Mata
``The photon-box Bohr-Einstein debate demithologized'' Eur. J.
Phys. \textbf{21}, 253-260 (2000).

\bibitem{grif}
David Griffiths, {\em Introduction to Elementary Particles}.
Harper and Row   (1987), (p.52).

\bibitem{ng3} W. A. Cristiansen, Y. J. Ng and H. van Dam,
``probing spacetime foam with extragalactic soources'' arXiv:
gr-qc/0508121
\end{thebibliography}
\end{document}